\documentclass[reqno]{amsart}

\usepackage{calrsfs}
\usepackage{verbatim}
\usepackage[utf8]{inputenc}
\usepackage[margin=2cm]{geometry}
\def\re#1{(\ref{#1})}

\newcommand\nc{\newcommand*} \nc\longnc{\newcommand}

\nc\emp\textit 
\nc\lat\textit 
\nc\ie{\lat{i.e.,\ }} \nc\etal{\lat{et al.\ }} \nc\etc{\lat{etc.\ }}
\nc\eg{\lat{e.g.,\ }} \nc\insitu{\lat{in situ}} \nc\QED{\lat{Q.E.D.}}
\nc\cf{cf.\ } \nc\wrt{w.r.t.\ }
\nc\lhs{l.h.s.\ }
\nc\rhs{r.h.s.\ }

\nc\e{\mathrm{e}} 
\nc\dd{\mathrm{d}} 
\nc\pd\partial
\nc\pdt{\pd_t}

\nc\qrho{\rho}

\nc\qlam{\lambda}

\begin{document}
\title{Open mathematical aspects of continuum thermodynamics: hyperbolicity, boundaries and nonlinearities}

\author{Mátyás Szücs$^{1,2}$, Róbert Kovács$^{1,2,3}$, Srboljub Simi\'c$^{4}$}

\address{%
$^{1}$  Department of Energy Engineering, Faculty of Mechanical Engineering, BME, 1521 Budapest, Hungary; \\
$^{2}$  Montavid Thermodynamic Research Group, 1112 Budapest, Hungary; \\
$^{3}$  Department of Theoretical Physics, Wigner Research Centre for
Physics, Institute for Particle and Nuclear Physics, 1525 Budapest, Hungary; \\
$^{4}$  Department of Mathematics and Informatics, Faculty of Sciences, University of Novi Sad, Novi Sad, Serbia}

\begin{abstract}Thermodynamics is continuously spreading in the engineering practice, which is especially true for non-equilibrium models in continuum problems. Although there are concepts and approaches beyond the classical knowledge, which are known for decades, their mathematical properties and consequences of the generalizations are less-known and are still of high interest in current researches. Therefore, we found it essential to collect the most important and still open mathematical questions related to different continuum thermodynamic approaches. First, we start with the example of Classical Irreversible Thermodynamics (CIT) in order to provide the basis for the more general and complex frameworks, such as the Non-Equilibrium Thermodynamics with Internal Variables (NET-IV) and Rational Extended Thermodynamics (RET). Here, we aim to present that each approach has its specific problems, such as how the initial and boundary conditions can be formulated, how the coefficients in the partial differential equations are connected to each other, and how it affects the appearance of nonlinearities. We present these properties and comparing the approach of NET-IV and RET to each other from these points of view.
\end{abstract}
\maketitle

\pagestyle{plain}

\section{Introduction}
The famous finding of Fourier, widely known as Fourier's law in which the heat flux $q_i$ is proportional with the temperature gradient $\partial_i T$,
\begin{align}
q_i=-\lambda \partial_i T, \label{Feq}
\end{align}
serves the basis for almost all the thermal problems in the engineering practice, where $\lambda$ is known to be the thermal conductivity. Here, and throughout the manuscript, we use index notation with Einstein's summation convention. However, as many experimental results prove, eq.~\eqref{Feq} must be generalized in order to provide a reliable model both for describing heat waves and room temperature phenomena \cite{Acketal66, McN74t, Botetal16, JozsKov20b}, occurring on a wide range of spatial scales. 
This is a task that must be carried out carefully, and in which thermodynamics has an inevitable role with increasing importance. 

Independently of how one generalizes eq.~\eqref{Feq}, its consequences must be understood before starting to choose and utilize a model in practical problems, beyond simply trying to model an experimental setting. This stands as a motivation for the present paper: discussing how the thermodynamic modeling process restricts the mathematical properties of the resulting partial differential equations, and how it aids (or fades) our understanding of a physical phenomenon. For instance, generalizing eq.~\eqref{Feq} modifies the possibilities for initial and boundary conditions, their analytical and numerical solutions, and their compatibility for coupled problems, i.e., what is a physically feasible, reasonable and solvable model. Therefore, one needs a framework that is constructive, easy to handle for practical problems but still provides insight that is enough for the problem.

In order to fulfill these requirements, many approaches are developed in the last decades. The framework of Classical Irreversible Thermodynamics (CIT) is well-understood for classical phenomena such as diffusive heat conduction or coupled problems (diffusive heat and moisture transport, Peltier and Seebeck effects, etc.) \cite{deGroot1962, Prigogine1947}. Even this classical approach of CIT contains some interesting and less-known possibilities for modeling. Section 2 is devoted to present the usual procedure of CIT, with discussing its mathematical consequences. Section 3 aims to continue and broaden the classical approach by introducing the so-called internal variables. This is the approach of Non-Equilibrium Thermodynamics with internal variables (NET-IV) \cite{Gyarmati1970, BerVan17b, JozsKov20b}. Here, we focus mainly on nonlinearities, initial and boundary conditions, and its connection with the framework of Rational Extended Thermodynamics (RET) \cite{RET, RETpoly}, b
 y discussing briefly an intermediate approach, called Extended Irreversible Thermodynamics (EIT) \cite{JouVasLeb88ext}. Both RET and EIT follow the results and procedure of kinetic theory, but on a different level, this is the reason why EIT is called to be intermediate between NET-IV and RET \cite{KovEtal18rg}. Therefore, Section 4 presents the basic structure of RET together with its mathematical consequences. More closely, we concentrated on the closure problem particulary, which affects the boundary conditions differently than in the case of NET-IV. In the sense of methodology and problems which are analyzed, this study complements recent review \cite{cimmelli2014entropy} and can be regarded as the other side of the same coin. 

\section{Classical Irreversible Thermodynamics}
Here, we aim to briefly summarize the classical approach in order to ease the understanding of the later Sections. 
In this regard, we refer to the work of Onsager, Eckart, de Groot-Mazur, Prigogine and Gyarmati \cite{Onsager1931a,Onsager1931b,Eckart1940a,Eckart1940b,Eckart1940c,Eckart1948, deGroot1962, Prigogine1947, Gyarmati1970},
who have founded the bases of irreversible thermodynamics in continuum physics. 
The importance of CIT lies in the following:
\begin{enumerate}
    \item to ensure a unified thermodynamical background for the classical transport equations, e.g., for Fourier heat conduction, Fick diffusion, Newton and Stokes viscosity laws;
    \item to discuss different cross-effects systematically, such as Soret-, Dufour-, Peltier-, and Seebeck-effects.
\end{enumerate}
In general, the continuum theories are formed by balance equations, constitutive equations, and equations of states. Here, in case of CIT, and later in NET-IV, the theory starts from the balances, which are `closed' by the constitutive relations. In RET, it is different, there the balances are obtained and closed differently. In the following, we restrict ourselves on a single component continuum. 
Table \ref{balances} summarizes the integral and the differential form of the balance equations for both the local and substantial descriptions, in which we followed Gyarmati \cite{Gyarmati1970} in notation: $ X $ denotes any extensive thermodynamical state quantity, and $ x := \frac{X}{M}$ is $ X $'s mass-specific quantity, where $ M $ is the mass of the body. Furthermore, $ \left( J_X \right)_i $ is the so-called substantial or conductive current density of $ X $, $ n_i $ is the normal vector of the differential surface element $ \dd A $, and $ \Sigma_X $ is the volumetric production rate of $ X $. Moreover, we distinguish the spatial domains and their boundaries accordingly with respect to the description: $ \mathcal{B} $ stands for the collection of points  comoving with the body, therefore it is not a fixed domain, contrary to $ \Omega $. Consequently, there is no mass flux through $ \pd \mathcal{B} $, but the mass flux on $ \pd \Omega $ contributes to the other balances
 .

\begin{table}[!h]
\centering
\begin{tabular}{c|cc}
& Material/Substantial description & Spatial/Local description \\
\hline \\
Integral form & $ \int\limits_{ \mathcal{ B } } \qrho \dot x \dd V = - \oint\limits_{ \pd \mathcal{ B } } \left( J_X \right)_i n_i \dd A + \int\limits_{ \mathcal{ B } } \Sigma_X \dd V $ & $ \int\limits_{ \Omega } \pdt \left( \qrho x \right) \dd V = - \oint\limits_{ \pd \Omega } \left[ \left( J_X \right)_i + \qrho x v_i \right] n_i \dd A + \int\limits_{ \Omega } \Sigma_X \dd V $ \\
Differential form & $ \qrho \dot x = - \pd_i \left( J_X \right)_i + \Sigma_X $ & $ \pdt \left( \qrho x \right) = - \pd_i \left[ \left( J_X \right)_i + \qrho x v_i \right] + \Sigma_X $
\end{tabular} 
\caption{Summarizing the balance equation} \label{balances}
\end{table}

From now on, we use the upper dot notation for the substantial time derivative, i.e., $ \dot{ \ } \equiv \frac{ \dd }{ \dd t } = \pd_t + v_i \pd_i $, which shows how a quantity changes in time at a fixed material point while it moves with velocity $ v_i $, and, $ \pdt $ and $ \pd_i $ are the partial time and space derivatives, respectively.

With omitting the details, the following substantial balance equations can be derived for the thermomechanical\footnote{Chemical, moisture diffusion and electromagnetic processes can also be included, but here we neglect them.} processes of non-polar\footnote{In non-polar continua, the internal rotational degrees of freedom play no role, thus the angular momentum is not an independent state variable.} continua:
\begin{enumerate}
    \item Conservation of mass implies the so-called continuity equation
    \begin{align}
        \label{cont}
        \dot \qrho + \qrho \pd_i v_i = 0 .
    \end{align}
    \item Cauchy's first equation of motion
    \begin{align}
        \label{Cauchy1}
        \qrho \dot v_i = \pd_j \sigma_{ij} + \qrho g_i
    \end{align}
    ensures the conservation of linear momentum. Here $ \sigma_{ij} $ denotes the Cauchy stress tensor, which is the opposite of the pressure tensor $ P_{ij} $ and $ g_i $ is the field strength.
    \item The conservation of angular momentum implies the symmetry of the Cauchy stress, i.e.,
    \begin{align}
        \label{Cauchy2}
        \sigma_{ij} = \sigma _{ji} ,
    \end{align}
    which is valid only for non-polar media, and usually called as Cauchy's second equation of motion.
    \item Starting with the conservation of total energy, one can derive the balance of internal energy or the first law of thermodynamics, which reads as
    \begin{align}
        \label{1stlaw}
        \qrho \dot e = - \pd_i q_i + \sigma_{ij} \pd_j v_i + q_V ,
    \end{align}
    where $ e $ denotes the specific internal energy, $ q_i $ the heat flux density (more precisely the conductive current of internal energy), $ q_V $ is the volumetric heat source density, $ \sigma_{ij} \pd_j v_i $ is the mechanical power, for which $  \sigma_{ij} \pd_j v_i = \sigma_{ij} \left( \pd_j v_i + \pd_i v_j \right) / 2 $ holds in the light of \re{Cauchy2}.
\end{enumerate}
These balance equations -- prescribed for conservative quantities -- are governed by the entropy balance, which is
\begin{align}
    \label{bal:subst-s}
    \qrho \dot s = - \pd_i \left( J_S \right)_i + \Sigma_S,
\end{align}
in substantial form, and $ s $ is the mass specific entropy, $ \left( J_S \right)_i $ is the entropy current density and $ \Sigma_S $ is the entropy production, which is non-negative and measures the dissipation along a process. 
While the first law of thermodynamics is given in \eqref{1stlaw}, the second law is more complex. Although, it is expressed formally by \eqref{bal:subst-s}, we provide here a more detailed definition, following Ván \cite{Van2018}.

\begin{enumerate}
    \item There exists independent themodynamical bodies, which can be described by extensive state quantities and  intensive state functions. One body is characterized by $ N $ extensive state quantities ($ X_a, \ a \in 1,2,\dots,N $). The  thermodynamical state space is spanned by the Descartes product of all $ X_a$. We will denote the intensive state functions with $ Y_a, \ a \in 1,2,\dots,N $, which are the functions of the extensive ones. For fluids and gases with one component, the extensive state quantities are the mass $ M $, the volume $ V $ and the internal energy $ E $.
    \item The entropy is the potential function of the vector space spanned by the intensive thermodynamical state quantites, characterized by the Gibbs relation as
    \begin{align}
        \dd S \left( X_1, X_2, \dots, X_N \right) = \sum_{ a = 1 }^{ N } Y_a \dd X_a .
    \end{align}
    In other words, the intensive state functions are the partial derivatives of entropy, i.e.,
    \begin{align}
        Y_a \left( X_1, X_2, \dots, X_N \right) &= \left. \frac{ \pd S }{ \pd X_a } \right|_{ X^b ,\ a \neq b } , &
        a,b &= 1,2,\dots,N .
    \end{align}
    For fluids and gases the Gibbs relation is
    \begin{align}
        \dd S ( E, V , m ) = \frac{ 1 }{ T } \dd E + \frac{ p }{ T } \dd V - \frac{ \mu }{ T } \dd M,
    \end{align}
    where the intensive state functions are defined by the partial derivatives
    \begin{align}
        \frac{ 1 }{ T } ( E, V , m ) &= \left. \frac{ \pd S }{ \pd E } \right|_{ V , M } , &
        \frac{ p }{ T } ( E, V , m ) &= \left. \frac{ \pd S }{ \pd V } \right|_{ E , M } , &
        - \frac{ \mu }{ T } ( E, V , m ) &= \left. \frac{ \pd S }{ \pd M } \right|_{ E , V } ,
    \end{align}
    in which $ T $ denotes the temperature, $ p $ is the (static) pressure and $ \mu $ is the chemical potential.
    \item The entropy is an extensive thermodynamical function, called as locality condition. More precisely,
    \begin{enumerate}
        \item the entropy is an Euler homogeneous function of any of its variables, e.g.,
        \begin{align}
            S \left( \qlam X_1, X_2, \dots, X_N \right) &= \qlam S \left( X_1, X_2, \dots, X_N \right) , & \qlam \in \mathbb{ R }^+ .
        \end{align}
        \item for all scalar $ X_A $, one can be introduce the $ X_A $-specific entropy, which is the function of the corresponding specific quantities, e.g., the $ X^1 $-specific entropy $ s $ is defined as
        \begin{align}
            S \left( X_1, X_2, \dots, X_N \right) = X_1 s \left( \frac{ X_2 }{ X_1 }, \dots, \frac{ X_N }{ X_1 } \right) .
        \end{align}
        Usually the mass specific entropy $ s ( e , v ) $ and the entropy density $ \qrho_S \left( \qrho_E , \qrho \right) $ is used, where $ e $ is the (mass) specific internal energy, $ v $ is the specific volume, and $ \qrho_E = \frac{ E }{ V } $ is the internal energy density.
        \item The {Euler relation} for entropy reads as
        \begin{align}
            \label{Ms}
            S \left( X_1, X_2, \dots, X_N \right) = \sum_{ a = 1 }^{ N } Y_a \left( X_1, X_2, \dots, X_N \right) X_a ,
        \end{align}
        which, for fluids and gases, is
        \begin{align}
            \label{S(E,V,M)}
            S ( E , V , M ) = \frac{ 1 }{ T } E + \frac{ p }{ T } V - \frac{ \mu }{ T } M .
        \end{align}
        In the light of \re{Ms}, the entropy can be given as
        \begin{align}
            \label{S=Ms}
            S ( E , V , M ) = M s ( e , v ) = M \left( \frac{ 1 }{ T } e + \frac{ p }{ T } v - \frac{ \mu }{ T } \right) ,
        \end{align}
        where the expression in the bracket is the mass specific entropy.
    \end{enumerate}
    As a consequence of statement 3, the Gibbs relation follows for $s ( e , v )$:
     \begin{align}
        \dd s = \frac{ 1 }{ T } \dd e + \frac{ p }{ T } \dd v .
    \end{align}
    \item The $ X_1 $-specific entropy is a concave function of its variables, i.e.,
    \begin{align}
        \det{ \pd_{ x_a , x_b } s ( x_2 , \dots, x_N )} & \le 0 , & a,b &= 2,\dots,N .
    \end{align}
    This statement leads to the internal or material stability criteria. Particularly for fluids and gases, these criteria are
    \begin{align}
        c_v &:= \left. \frac{ \pd e ( T , v ) }{ \pd T } \right|_{ v } > 0 , &
        \chi_T &:= - \frac{ 1 }{ v } \left. \frac{ \pd p ( T , v ) }{ \pd v } \right|_{ T } > 0 ,
    \end{align}
    where $ c_v $ is the isochoric specific heat and $ \chi_T $ is the isothermal compressibility.
\end{enumerate}

All these together form the concept of entropy, and the second law is expressed through its balance equation. 
Moreover, CIT has one more crucial assumption, which is called local equilibrium hypothesis.
Usually, the classical literature defines it as the state of a material point is close to the equilibrium, that is, all the spatial points of a continuum is close to equilibrium, and quasi-equilibrium states are following each other from time instant to time instant. Such definition suffers from many ambiguities, e.g., how the notions `close', and `quasi-equilibrium' can be defined. Instead, we recommend using the following definition.
Local equilibrium is when the Gibbs-relation remains valid out of homogeneous equilibrium. In other words, the same $X_a$ extensives are used to characterize both the equilibrium state and the process out of equilibrium. 

However, it is clear by now that the local equilibrium hypothesis has a limited region of validity. For instance, heterogeneous materials, fast processes, very high heat fluxes, low or high-temperature conditions, etc., are all capable of resulting in a non-equilibrium phenomenon. Consequently, in the generalizations of CIT, other variables are introduced as well, called non-equilibrium variables. 

\subsection{Exploiting the second law}

Equations \re{cont}, \re{Cauchy1} and \re{1stlaw} determine the time evolution of mass, momentum and energy fields, and containing five unknown quantities ($ \qrho $, $ v_i $, $ \sigma_{ij} $, $ e $, $ q_i $, but $ g_i $ and $ q_V $ are prescribed functions) overall. From mathematical poing of view, one needs a `closure', the relation between densities and their fluxes. These equations are the so-called constitutive equations, (roughly, the material laws), which describe the relationships among $ \left( J_X \right)_i $ and $ x $. 

Thermodynamically consistent constitutive equations can be derived with the help of entropy and entropy production. The reversible---or the entropy preserving---contribution of the constituve equations is embedded into the entropy function, while the irreversible---or dissipative---contribution, which increases entropy (thus results in thermodynamically stable processes) can be derived via evaluating the entropy production. For mathematically rigorous techniques, we refer to the work of Colemann--Noll and Liu \cite{Coleman1963, Liu1972}.

In what follows, we provide a simple example by utilizing a mathematically less precise, but and `easier to understand and use' method. 

\noindent
\textbf{Example: Heat conduction in rigid bodies -- Fourier heat conduction.}
Here, we consider a rigid body at rest \wrt a given reference frame, thus both the velocity field and the volume change are identically zero. Furthermore, we neglect the volumetric heat source density, too. Consequently, the balance of internal energy and the Gibbs relation are
\begin{align}
    \label{bal:e-hc}
    \qrho \dot e = - \pd_i q_i; \quad \dd s &= \frac{ 1 }{ T } \dd e + \underbrace{ \frac{ p }{ T } \dd v }_{ = 0 }.
\end{align}
Let us evaluate now the \lhs of the entropy balance \re{bal:subst-s}, constrained by the balance of internal energy \re{bal:e-hc}. With separating a full divergence term, we have
\begin{align}
    \qrho \dot s \stackrel{ \re{bal:e-hc} }{ = } \frac{ 1 }{ T } \qrho \dot e \stackrel{ \re{bal:e-hc} }{ = } - \frac{ 1 }{ T } \pd_i q_i \stackrel{ \re{bal:subst-s} }{ = } - \pd_i \Big( \underbrace{ \frac{ 1 }{ T } q_i }_{ = \left( J_S \right)_i } \Big) + \underbrace{ q_i \pd_i \frac{ 1 }{ T } }_{ = \Sigma_S } .
\end{align}
First, by separating the full divergence from $\qrho \dot s$, we obtained the usual definition of the entropy current density, i.e., it is not necessarily an apriori assumption on contrary to the usual approach, and reads
\begin{align}
    \label{J_S}
    \left( J_S \right)_i = \frac{ 1 }{ T } q_i.
\end{align}
In case of the generalized theories, \eqref{J_S} is different and offers the possibility to include spatially nonlocal effects in the model (roughly speaking, higher-order spatial derivatives).
Finally, one must ensure the positive semi-definiteness of the entropy production
\begin{align}
    \Sigma_S = q_i \pd_i \frac{ 1 }{ T } ,
\end{align}
for which the simplest solution is to assume a linear dependence on reciprocal temperature gradient,
\begin{align}
    q_i = l \left( T \right) \pd_i \frac{ 1 }{ T }= - \frac{ l ( T ) }{ T^2 } \pd_i T = - \qlam ( T ) \pd_i T ,
\end{align}
where the heat current density appears to be as a linear function of $\pd_i \frac{ 1 }{ T }$. Moreover, $ l \left( T \right) $ is a positive function of temperature, and $ \qlam ( T ) $ is called thermal conductivity. There are several -- infinite -- other opportunities to ensure the positive-semidefiniteness of entropy production. For example,
\begin{align}
    q_i = l_1 ( T ) \pd_i \frac{ 1 }{ T } + l_3 ( T ) \left( \pd_i \frac{ 1 }{ T } \right)^3 + l_5 ( T ) \left( \pd_i \frac{ 1 }{ T } \right)^5 + \dots ,
\end{align}
where $ l_1 ( T ) $, $ l_3 ( T ) $, $ l_5 ( T ) $ are positive functions, and the positiveness of $\Sigma_S$ is immediately ensured. {Analogously, assuming $l(T, \partial_i T)$ coefficient results in a similar solution. It is still unknown how these nonlinear solutions could relate to the experiments.}

Overall, heat conduction in a rigid body can be described by the balance of internal energy, and the constitutive equation, called Fourier's law in this case:
\begin{align}
    \label{bal:e-hc-1}
    \qrho \dot e &= - \pd_i q_i , \\
    \label{Fourier}
     q_i &= - \qlam ( T ) \pd_i T .
\end{align}
As we mentioned above, the temperature is a function of the internal energy. Assuming that $ T(e) $ is invertable, and the specific internal energy is linear in the temperature with the specific heat $ c $ as $ e (T) = c T $, then substituting \re{Fourier} into \re{bal:e-hc-1}, the heat conduction equation for temperature is
\begin{align}
    \qrho c \dot T &=  \pd_i \left[ \qlam ( T ) \pd_i T \right] ,
\end{align}
which is a parabolic second order partial differential equation. From mathematical point of view, the usual boundary conditions are the following. 
\begin{enumerate}
    \item First kind or Dirichlet boundary condition: the temperature is prescribed on the boundary:
    \begin{align}
        \label{BC1}
        T ( t , r_i )|_{ \pd \mathcal{B} } = T_{ \pd \mathcal{B} } ( t ) ,
    \end{align}
    here $ t $ and $ r_i $ are the time and space coordinates, respectively.
    \item Second kind or Neumann boundary condition: the normal component of the heat flux is prescribed on the boundary, that is,
    \begin{align}
        \label{BC2}
        \left( q_i ( t , r_i ) n_i \right)|_{ \pd \mathcal{B} } = q_{ \pd \mathcal{B} } ( t ).
    \end{align}
    Applying the Fourier equation \re{Fourier}, \eqref{BC2} can be rewritten to
    \begin{align}
        \left( \pd_i T ( t , r_i ) n _i \right) |_{ \pd \mathcal{B} } = - \frac{ 1 }{ \qlam }q_{ \pd \mathcal{B} } ( t ) .
    \end{align}
    \item Third kind or Robin boundary condition (convection type): the linear combination of the temperature and the normal component of heat flux is prescribed on the boundary. For instance, the Newtonian cooling boundary is
    \begin{align}
        \left( q_i ( t , r_i ) n_i \right)|_{ \pd \mathcal{B} } = h \left[ T ( t , r_i )|_{ \pd \mathcal{B} } - T_{ \infty } \right] ,
    \end{align}
    where $ h $ is the heat transfer coefficient and $ T_{ \infty } $ is the ambient temperature.
\end{enumerate}

Here we add some comments on the boundary conditions. Let us assume, that the material parameters $ c $ and $ \qlam $ can be considered as constant. In this case one can introduce the thermal diffusivity $ \alpha := \frac{ \qlam }{ \qrho c } $, and the heat conduction equation reduces to
\begin{align}
    \dot T = \alpha \pd_i \pd_i T .
\end{align}
However, since the differential operator $ \dot{ \ } - \alpha \pd_{i,i} $ is linear, a similar equation can be derived for the heat current density as well,
\begin{align}
    \dot q_i = \alpha \pd_i \pd_k q_k .
\end{align}
in which $q_i$ stands as the only field variable. These can be called as the temperature and heat flux representations of the system \eqref{bal:e-hc-1}-\eqref{Fourier}. 
For some particular situations, such as the modeling of the laser flash experiment, $q$-representation fits well. Notably, in this case, the usual second kind boundary becomes first kind, and it is not possible anymore to prescribe the temperature directly on the boundary.
However, knowing the $q_i$ field, the temperature can be easily recovered by integration, with a further restriction on a reference point.

Similarly, the constitutive equations for heat conducting Newtonian fluids, i.e., the Navier-Stokes-Fourier equations is also possible to derive in the same way \cite{JozsKov20b}.

\section{Non-Equilibrium Thermodynamics with Internal Variables} 
While the classical thermal and mechanical equations - Hooke, Navier-Stokes, Fourier - are enough for the majority of engineering problems, these models have their limitations as any other model as well. For instance, modeling rheological phenomena occurring in solids require the generalization of Hooke's law. One possible way to do so is to introduce dynamic degrees of freedom \cite{Verhas97}, aimed to catch the strain rate dependence by including a non-equilibrium state variable. A more general approach is to use internal variables, which are not necessarily zero in equilibrium, contrary to dynamic degrees of freedom. This is more apparent in situations related to microcrack modeling, in which the density function of cracks can be non-zero in equilibrium \cite{VanEtal00}. In the following, we place our focus on internal variables through the example of the Maxwell-Cattaneo-Vernotte (MCV) and Guyer-Krumhansl (GK) equations \cite{Max1867, Cattaneo58, Vernotte58, GuyKru66a1, GuyKru66a2}. 

\subsection{Maxwell-Cattaneo-Vernotte equation}
In both cases, we follow the usual procedure of non-equilibrium thermodynamics \cite{BerVan17b, JozsKov20b}, in which the classical, local equilibrium part of the specific entropy $s$ is shifted by a positive semidefinit function of the internal variable $\xi_i$ in a way to preserve the convexity properties of the entropy:
\begin{align}
s(e,\xi_i)=s_e(e) - \frac{m}{2}\xi^2, \quad (m \in \mathbb R^+_0). \label{smcv}
\end{align}
As a special choice, one can decide to identify $\xi_i$ with $q_i$, and thus achieve compatbility with EIT. Applying the classical entropy flux $(J_S)_i=q_i/T$, one obtains the following expression for the entropy production with considering $\xi_i \equiv q_i$:
\begin{align}
\Sigma_s=\rho \dot s(e,\xi_i) + \partial_i (J_S)_i = q_i \left ( \partial_i \frac{1}{T} - \rho m \dot q_i \right ) \geq 0, \label{epMCV}
\end{align}
and the balance of internal energy,
\begin{align}
\rho \dot e + \partial_i q_i =0
\end{align}
is exploited, in which the source terms are neglected. Here we encounter multiple possibilities how to solve the entropy inequality \eqref{epMCV}, similarly to the Fourier equation. First, seeking linear solutions for isotropic materials, we find the following
\begin{align}
q_i = l \left ( \partial_i \frac{1}{T} - \rho m \dot q_i \right ), \quad (l \in \mathbb R^+), \label{MCVbf}
\end{align}
in which one can define the thermal conductivity $\lambda = l/T^2$ and the relaxation time $\tau=\rho m l$, and obtain the final form of the MCV equation as
\begin{align}
\tau \dot q_i + q_i = - \lambda \partial_i T. \label{MCVeq}
\end{align}
Although this derivation is well-known since Gyarmati \cite{Gyar77a}, there are multiple mathematical aspects, which must be discussed in detail. 
First, the Eq.~\eqref{MCVbf} shows a sort of `basic form' of the constitutive equation, where the relationship among the coefficients is visible. It is of high importance when the simplest nonlinearities, e.g., temperature-dependent thermal conductivity is tried to be implemented: it immediately affects the relaxation time as well with far-reaching consequences \cite{KovRog20}. Therefore, it is not possible to arbitrarily include such nonlinearities without knowing the `basic form' of a constitutive equation. It is more apparent for the Guyer-Krumhansl equation. Furthermore, since $q_i$ became a state variable, it is thermodynamically possible to assume $l=l(T,q_i)$, which case is entirely open for future investigation both theoretically and experimentally. {An analogous but distinct approach is also published recently \cite{DomEtal20}, in which the nonlinearities are treated on a different ground with a similar $q_i$ dependence.}

Second, Eq.~\eqref{MCVbf} is only the simplest linear solution, and the entropy inequality enables to find infinitely many nonlinear ones, too. For instance, the `simplest' nonlinear solution arises as a polynomial of $\left ( \partial_i \frac{1}{T} - \rho m \dot q_i \right )$. However, even the linear case is able to bring up difficulties, still open questions both on physical and mathematical grounds. 
Such a difficulty emerges when non-uniform initial temperature distribution is accounted. In case of Fourier's law, it is quite straightforward how to handle that situation: the gradient of the initial temperature field immediately leads to the initial state of the heat flux field. In case of the MCV equation, one has to solve the differential equation \eqref{MCVeq} first, together with a constraint to determine the integration coefficient. It seems to be `trivial', but let us keep in mind that the MCV equation is the simplest extension of the Fourier equation with merely one time derivative. 

\subsection{Guyer-Krumhansl equation}
The situation becomes more difficult for the GK equation. Its derivation requires a generalized entropy flux as well, beyond the specific entropy \eqref{smcv}. In the framework of EIT, $(J_S)_i=q_i/T + \mu \partial_iq_j q_j$, $(\mu \in \mathbb R^+_0)$ is assumed, which can exactly be recovered in the approach of NET-IV using a current multiplier, or Nyíri-multiplier $B_{ij}$ \cite{Nyiri91}. Here we have two options: either $(J_S)_j= B_{ij} q_i$ or $(J_S)_j=q_j/T+B_{ij} q_i$ are valid. 
Calculating the entropy production, for the first option we obtain
\begin{align}
\Sigma_s = \partial_j q_i \left ( B_{ij} - \frac{1}{T} \delta_{ij} \right ) + q_i (\partial_j B_{ij} - \rho m \dot q_i) \geq 0,
\end{align}
and for the second one, it reads
\begin{align}
\Sigma_s = \partial_j q_i  B_{ij} + q_i \left (\partial_j B_{ij} + \partial_i\frac{1}{T} - \rho m \dot q_i \right ) \geq 0.
\end{align}
In both cases, $B_{ij}$ is defined through the entropy inequality, and can recover the EIT approach exactly by assuming the simplest linear solution. We note that both cases allow a more general solution for $B_{ij}$, e.g., $B_{ij} = \mu_{ijkl}\partial_lq_k$ in which case  $\mu_{ijkl}$ is a fourth order tensor. 

In order to keep the discussion focused, let us restrict ourselves to the one-dimensional situation and find the following as a solution of the entropy inequality:
\begin{align}
B - \frac{1}{T} &= l_1 \partial_x q,  (l_1 \in \mathbb R_0^+) \\
\partial_x B - \rho m \dot q &=l_2 q, \quad  (l_2 \in \mathbb R^+)
\end{align}
which forms the GK equation:
\begin{align}
\tau \dot q + q = - \lambda \partial_x T + \kappa^2 \partial_{xx} q, \label{GKeq1D}
\end{align}
where
\begin{align}
\tau = \frac{\rho m}{l_2}, \quad \lambda = \frac{1}{l_2 T^2}, \quad \kappa^2 = \frac{l_1}{l_2}.
\end{align}
Notably, the Nyíri-multiplier is an efficient tool to realize coupling between quantities with different tensorial order, which would not be possible otherwise for isotropic materials due to the Curie principle. Physically, the spatially nonlocal extension of a hyperbolic model, i.e., the appearance of new spatial derivatives in the constitutive equation comparing to the MCV \eqref{MCVeq}, leads to a parabolic model \cite{JozsKov20b}. Consequently, a Nyíri-multiplier ensures a parabolic envelope for hyperbolic equations. Also, in general the occurrence of the nonlocal behaviour implies a coupling between different tensorial order quantities. Here, it can be interpreted as a coupling between $q_i$ and $\partial_j q_i$, which $\partial_j q_i$ represents the rate of a `thermal pressure' \cite{KovVan15}. 

Regarding the state variable dependent coefficients, the importance of the `basic form' is more apparent here. For instance, $l_2=l_2(T) : \mathbb R^+ \rightarrow \mathbb R^+$, $(l_2\in \mathcal C^1)$ introduces the T-dependence into all other coefficients. It is shown in \cite{KovRog20} on the example of the MCV equation that achieving linear T-dependence in $\tau$ and $\lambda$ necessarily implies $\rho=\rho(T)$, therefore the mechanics must be included. Although $\lambda=\lambda(T)$ looks completely natural in the classical case, for non-Fourier problems this convenience vanishes.
 Moreover, if $l_1=l_1(T): \mathbb R^+ \rightarrow \mathbb R^+$, $(l_1\in \mathcal C^1)$ holds, one encounters extra derivatives in \eqref{GKeq1D}, e.g.,
\begin{align}
\frac{\textrm{d}l_1}{\textrm{d}T}\partial_x T \partial_x q + l_1(T) \partial_{xx} q + \partial_x \frac{1}{T} - \rho m \dot q=l_2 q,
\end{align}
which seems to be straightforward if the GK equation is given in its `basic form'; however, this is not the case in the literature, in general. Therefore we suggest that speaking about non-Fourier models is safe only in the case when the complete picture behind the constitutive equation is provided, especially for nonlinear problems. Otherwise, the implementation of any state variable dependence is unwise and implies unkown simplifications as well.

The treatment of non-uniform initial conditions becomes significantly difficult contrary to Fourier's law. Since the constitutive equation \eqref{GKeq1D} consists of both time and spatial dependence, one must include the boundary conditions in order to determine the initial heat flux distribution, too. This is unusual in the classical approach. Furthermore, even the initially uniform (hence $\partial_i T = 0_i$) temperature distribution could allow a more general, possibly non-uniform heat flux distribution. The only requirement is to satisfy the following partial differential equation at the initial time instant:
\begin{align}
\tau \dot q + q = \kappa^2 \partial_{xx} q.
\end{align}
The trivial solution is a constant $q$ at $t=0$, enabling the non-zero case, too. Now, it reveals the next obstacle to applying the classical approach for boundary conditions. For the Fourier equation, $q_i$ and $T$ are connected by the equality \eqref{Feq}, thus both field variables are adequate to define the usual boundary conditions. This does not hold for the GK equation, the nonlocal term $\partial_{xx} q$ does not allow the parallel prescription of boundaries for both variables \cite{RietEtal18, Kov18gk}. This is the primary disadvantage of finite element methods in which all field variables are placed onto the same node, hence producing false solutions \cite{FulEtal20}. This attribute of the GK equation can be handled in two ways. 

The first one is more suitable for analytical solutions; choose a primary field variable, which is used to prescribe the boundary conditions. Then, starting from the `basic form', it is possible to eliminate the other field variable, at least in the linear regime. This can be either $T$ or $q$-representation of the GK equation. It could also be  advantageous for the Fourier heat equation. 
The second option suggests keeping the evolution equations `untouched', i.e., in the form of a system of partial differential equations. When the evolution equations are thermodynamically compatible, it is possible to use a numerical method with staggered fields \cite{RietEtal18}, in which the primary field variable is kept on the boundary, while the other is shifted. Consequently, the shifted variable does not require to define boundary conditions, avoiding the problematic parallel description, and the structure of the evolution equations ensures the applicability of staggered fields. 
As a consequence of the proper treatment of boundary conditions, the solutions will satisfy the maximum principle, therefore the unphysical ones are excluded. This is not the case in \cite{Zhukov16}, where the temperature evolution presents negative values as well. 

\subsection{Further remarks about boundary conditions}

Finally, we mention a few general remarks about the formulation of boundary conditions. Let us recall the Robin-type boundary,
    \begin{align}
        \left( q_i ( t , r_i ) n_i \right)|_{ \pd \mathcal{B} } = h \left[ T ( t , r_i )|_{ \pd \mathcal{B} } - T_{ \infty } \right].
    \end{align}
In fact, the Fourier's law is its differential analogue, and that type of boundary represents the interaction between a point-like body and its environment through a thermal resistance. However, in principal, exchanging the constitutive equation would affect that definition, too. For example, considering the MCV equation, analogously the convective type boundary would be
\begin{align}
        \left(\tau \dot q_i ( t , r_i ) n_i  +  q_i ( t , r_i ) n_i \right)|_{ \pd \mathcal{B} } = h \left[ T ( t , r_i )|_{ \pd \mathcal{B} } - T_{ \infty } \right] ,
    \end{align}
One could argue for and against, but our aim here is to reveal that it is still an open, unsolved problem for non-Fourier models. This idea is already revealed in \cite{AlvJou10} for heat conduction models based on kinetic theory. However, the experimental verification is still missing, and based on \cite{AlvJou10}, the differences could be negligible in certain cases, but overall, there are no decisive results. { For a more detailed discussion about the boundary condition possibilities in internal variable theories, we refer to the work of Cimmelli \cite{Cimm02}.}

\subsection{Coupled heat and mass transport}
Previously, we focused our discussion on the structure of heat conduction equations. The same obstacles emerge for coupled systems, too; however, that parts serves as a bridge between NET-IV and RET. As the simplest demonstration for generalized coupled equations, we take the example of the extended Navier-Stokes-Fourier (NSF) equations. Their detailed thermodynamic background can be found in \cite{KovEtal18rg}, presenting their basis for approaches of NET-IV and RET. Briefly, the classical, local equilibrium part of the specific entropy $s(e,\rho)$ (with $\rho$ being the mass density) is extended in the same way as we have seen on the example of heat conduction, but now the mass density also plays an essential role. Thus the specific entropy for the generalized case is
\begin{align}
s(e,\rho,q_i, \Pi_{ij})= s_e(e,\rho) - \frac{m_1}{2} q_i q_i - \frac{m_2}{2} \Pi_{\langle ij \rangle} \Pi_{\langle ij \rangle} - \frac{m_3}{6} \Pi_{ii} \Pi_{jj}, \quad (m_1, m_2, m_3 \in \mathbb R_0^+),
\end{align}
in which $\Pi_{ij}$ is the viscous pressure, $\langle \ \rangle$ denotes its deviatoric part. Since the spherical and deviatoric parts are linearly independent of each other, that separation must be realized from the beginning, and utilized for the Nyíri-multiplier in the entropy current as follows:
\begin{align}
(J_S)_j=\left (B_{\langle ij\rangle} + \frac{B_{kk}}{3} \delta_{ij} \right)q_i.
\end{align}
Omitting the detailed calculation of entropy production $\sigma_s$, we arrive to 
\begin{eqnarray}
-\rho m_1 \dot q_i +\frac{1}{3} \partial_i b_{kk}  + \partial_j b_{\langle ji \rangle }&=& n q_i, \\
-\frac{1}{T} \partial_{\langle i} v_{j \rangle }- \rho m_2 \dot \Pi_{\langle ij \rangle}  &=& l_{11}\Pi_{\langle ij \rangle} + l_{12} \partial_{\langle i} q_{j \rangle }, \\
b_{\langle i j \rangle } &=&l_{21} \Pi_{\langle ij \rangle} + l_{22} \partial_{\langle i} q_{j \rangle }, \\
- \frac{1}{T} \partial_j v_j-\rho m_3 \dot \Pi_{ii} &=& k_{11} \frac{\Pi_{ii} }{3} + k_{12} \partial_i q_i, \\
 b_{kk} - \frac{1}{T} &=& k_{21} \frac{\Pi_{ii} }{3} + k_{22} \partial_i q_i
\end{eqnarray} 
with
\begin{align}
n>0, \quad l_{11} l_{22} - (l_{12}+ l_{21})^2/4 >0, \quad  k_{11} k_{22} - (k_{12} +k_{21})^2/4>0, \label{eNSFc0}
\end{align}
as the simplest linear solution for isotropic materials. Rearringing the coefficients with considering the strictly linear case, it reads
\begin{align}\label{eNSFeq}
\tau_1\dot q_i +q_i +\lambda \partial_i T - \alpha_{21} \partial_i \Pi_{kk} - \beta_{21} \partial_j \Pi_{\langle ij \rangle} & = \gamma_{1} \partial_{i} \left( \partial_{k} q_{k} \right)
+ \gamma_{2} \partial_{j} \left( \partial_{\langle i} q_{j \rangle} \right), \nonumber \\
\tau_2 \dot \Pi_{\langle ij \rangle} +\Pi_{\langle ij \rangle} + \nu \partial_{\langle i} v_{j \rangle} + \beta_{12} \partial_{\langle i} q_{j \rangle} &= 0, \\
\tau_3 \dot \Pi_{ii} +\Pi_{ii} + \eta \partial_i v_i + \alpha_{12} \partial_i q_i &= 0, \nonumber
\end{align}
where the coefficients are formed as
\begin{gather}
\tau_1 = \frac{\rho m_1}{n}, \quad \tau_2 = \frac{\rho m_2}{l_{11}}, \quad \tau_3=\frac{3 \rho m_3}{k_{11}}, \quad
\lambda=\frac{1}{3 n T^2}, \quad \nu=\frac{1}{T l_{11}},\quad \eta=\frac{3}{T k_{11}}, \nonumber \\
\alpha_{12}=\frac{3 k_{12}}{k_{11}} \quad \alpha_{21}=\frac{k_{21}}{9n}, \quad \beta_{12}=\frac{l_{12}}{l_{11}}, \quad \beta_{21}=\frac{l_{21}}{n}, \quad
\gamma_{1} = \frac{k_{22}}{3 n}, \quad \gamma_{2} = \frac{l_{22}}{n}; \label{eNSFc}
\end{gather}
$\tau_{1,2,3}$ are the corresponding relaxation times, $\nu$ and $\eta$ are the shear and bulk viscosities, respectively; the $\alpha$ and $\beta$ are the coupling coefficients.
We observe again that the appearance of the nonlocal terms necessarily implies a connection between different tensorial order quantities. While the utilization of Nyíri-multipliers allows to achieve such a coupling both for the deviatoric and spherical parts, it also results in a parabolic model \eqref{eNSFeq}. This parabolic structure includes further possibilities than its hyperbolic counterpart from RET (i.e., when $k_{22}=0, l_{22} = 0$). Consequently, the usual procedure of NET-IV leads to a more general model in which the coefficients are connected by means of \eqref{eNSFc0} and \eqref{eNSFc} without any further constraint, on the contrary to RET. In parallel, it could stand as a disadvantage as well due to the higher number of parameters to fit. As a side note, we may draw attention to a recent work \cite{rana2018coupled} which retains the local equilibrium assumption, and thus Gibbs relation, but allows for higher order coupling for entropy current similar to the one obtained by Nyíri-multipliers. As a result, coupled constitutive relations are obtained, but not of the rate type.

Staying in the linear and quasi-linear regime, the compatibility between RET and NET-IV (consequently, with EIT, too) is proved in \cite{KovEtal18rg}. However, the situation could be different in the nonlinear case, which becomes apparent regarding the rarefied gas experiments \cite{Kov18rg}. The importance of these experiments lies in the mass density dependence of the coefficients \eqref{eNSFc}. It is revealed in these experiments that the speed of sound considerably changes with respect to the mass density, keeping the temperature and the sound frequency $\omega$ constant. According to the original measurements, the outcome is the function of the frequency - pressure ratio of $\omega/p$ only. However, when the $\omega$ and $T$ are constants, the density remains a mere variable to change in order to reveal the change in the speed of sound. That is, decreasing the density of gas from its normal state down to the low-pressure region (e.g., $10$ Pa \cite{Kov18rg}), continuousl
 y enhance the role of the coupling between the fluid and thermal equations together with their hyperbolic character (i.e., the relaxation times become higher and higher). Therefore, these coefficients must be mass density-dependent. Indeed, it turned out to be true and crucial. 

On one hand, in order to recover the $\omega/p$ scaling in the dispersion relations, it demands a particular mapping, i.e., $z(\rho): \mathbb R^+ \rightarrow \mathbb R^+$ where $z$ could be any coefficient from \eqref{eNSFc}. Moreover, this scaling also requires constant viscosities, thermal conductivity, and ideal gas state equations. On the other hand, these requirements are contradictory with viscosity vs.~mass density measurements, in which the viscosity changes considerably for both the dense and rarefied regions. While models based on kinetic theory restrict themselves to such a scaling, the freedom in NET-IV offers further possibilities; for instance, it is easily possible to implement any measured mass density function with neither reaching further obstacles nor contradicting the basic principles of NET-IV.

Overall, we emphasize again that the implementation of nonlinearities requires special attention as the behavior of the generalized constitutive equations differs from the classical ones.

\section{Rational Extended Thermodynamics}
Extended thermodynamics emerged in the sixties of the last century \cite{muller1967paradox} as a method to overcome the paradox of infinite speed of pulse propagation, which appears in the Fourier model of heat conduction and propagation of shear diffusion in Newtonian fluids. Although these models are widespread in scientific literature, as well as engineering applications, their physical inconsistency with fundamental physical constraints seemed disturbing. 

The solution to this problem was first found within the framework of thermodynamics of irreversible processes (TIP), and therefore called Extended TIP. Further study related it more closely, in physical aspects, to the kinetic theory of gases, especially Grad's method of moments \cite{grad1949moments}, and to the hyperbolic systems of balance laws, in mathematical aspects. This direction is called Rational Extended Thermodynamics (RET), capturing at the same time methods and spirits of rational and extended thermodynamics \cite{RET}. 

It is clear from the outset that RET relies both on sound physical principles and desirable formal mathematical structure with the aim to establish physically/thermodynamically consistent models that predict a finite speed of pulse propagation. To that end, we shall discuss both aspects of this theory in the sequel. 

The section will be organized as follows. First, basic principles and formal structure of the governing equations will be presented, along with important mathematical consequences like symmetrization. Second, an alternative approach called Molecular Extended Thermodynamics (MET), based upon the application of the Maximum Entropy Principle (MEP) will be presented. Since the major issue in non-equilibrium thermodynamics is the closure problem, we shall demonstrate how these two approaches in conjunction lead to proper and complete closure of the governing equations. Finally, we shall tackle the important problem of boundary conditions. 

\subsection{Formalism of RET} 

Rational extended thermodynamics, as its name tells, has a rational foundation with twofold meaning of the adjective: on one hand, it is based upon physical principles; on the other, it constructs appropriate physical theory by deduction. Physical principles which serve as its basis are: 
\begin{enumerate}
	\item governing equations are of balance type; they are also invariant with respect to Galilean transformations;  
	\item constitutive functions depend on local (in point) values of field variables; 
	\item governing equations are adjoined with the entropy balance law with convex entropy density and non-negative entropy production. 
\end{enumerate}
The balance laws structure provides natural weak formulation and possibility for shock waves. Galilean invariance \cite{ruggeri1989galilean} determines the velocity dependence of constitutive quantities. The local character of the constitutive functions restricts the balance laws to the quasi-linear system of first-order partial differential equations. Finally, entropy inequality secures physical consistency, but also provides a tool for the transformation of governing equations into symmetric hyperbolic form. 

If we denote by $\mathbf{u}(t,\mathbf{x}) \in \mathbb{R}^{n}$ the vector of field variables, then generic form of governing equations in RET is: 
\begin{equation} \label{RET:BLaws}
  \partial_{t} \mathbf{F}(\mathbf{u}) + \partial_{k} \mathbf{F}_{k}(\mathbf{u}) 
    = \mathbf{f}(\mathbf{u}); \quad 
    \mathbf{F}_{k}(\mathbf{u}) = \mathbf{F}(\mathbf{u}) v_{k} 
    + \boldsymbol{\Phi}_{k}(\mathbf{u}),
\end{equation}
where $\mathbf{F}(\mathbf{u}) \in \mathbb{R}^{n}$ is vector of densities, $\mathbf{F}_{k}(\mathbf{u}) \in \mathbb{R}^{n}$ is vector of fluxes, $\boldsymbol{\Phi}_{k}(\mathbf{u}) \in \mathbb{R}^{n}$ is vector of non-convective fluxes, and $\mathbf{f}(\mathbf{u}) \in \mathbb{R}^{n}$ is vector of productions. Every solution $\mathbf{u}(t,\mathbf{x})$ of the system \eqref{RET:BLaws} is called thermodynamic process. 

If the vector of field variables is split into velocity $\mathbf{v} \in \mathbb{R}^{3}$ and vector of objective quantities $\mathbf{w} \in \mathbb{R}^{n-3}$, $\mathbf{u} = (\mathbf{v}, \mathbf{w})$, then invariance with respect to Galilean transformations restricts the velocity dependence of densities, fluxes and productions by means of matrix function $\mathbf{X}(\mathbf{v})$: 
\begin{equation} \label{RET:Galileo}
  \mathbf{F}(\mathbf{v}, \mathbf{w}) 
    = \mathbf{X}(\mathbf{v}) \hat{\mathbf{F}}(\mathbf{w}); \quad 
  \boldsymbol{\Phi}_{k}(\mathbf{v}, \mathbf{w}) 
    = \mathbf{X}(\mathbf{v}) \hat{\boldsymbol{\Phi}}_{k}(\mathbf{w}); \quad 
  \mathbf{f}(\mathbf{v}, \mathbf{w}) 
    = \mathbf{X}(\mathbf{v}) \hat{\mathbf{f}}(\mathbf{w}), 
\end{equation}
where hat denotes velocity independent functions determined by: 
\begin{equation*}
  \hat{\mathbf{F}}(\mathbf{w}) = \mathbf{F}(\mathbf{0}, \mathbf{w}); \quad
  \hat{\boldsymbol{\Phi}}_{k}(\mathbf{w}) 
    = \mathbf{\boldsymbol{\Phi}}_{k}(\mathbf{0}, \mathbf{w}); \quad
  \hat{\mathbf{f}}(\mathbf{w}) = \mathbf{f}(\mathbf{0}, \mathbf{w}), 
\end{equation*}
where $\mathbf{X}(\mathbf{0}) = \mathbf{I}$. 

Entropy inequality is a supplementary balance law that reads: 
\begin{equation} \label{RET:Entropy-BLaw}
  \partial_{t} h + \partial_{k} h_{k} = \Sigma \geq 0,  
\end{equation}
where $h$ is entropy density, $h_{k}$ is entropy flux, and $\Sigma$ is entropy production. Since \eqref{RET:Entropy-BLaw}, like balance laws \eqref{RET:BLaws}, must be invariant with respect to Galilean transformations, we have: 
\begin{equation*}
  h = h(\mathbf{w}), \quad h_{k} = h_{k}(\mathbf{w}), \quad \Sigma = \Sigma(\mathbf{w}). 
\end{equation*}
It restricts the structure of balance laws in more fundamental way: it must be satisfied for any thermodynamic process, i.e. for any solution of \eqref{RET:BLaws}. Since \eqref{RET:BLaws} and \eqref{RET:Entropy-BLaw} are both quasi-linear partial differential equations, they must be compatible. There comes entropy principle into play. In procedure established by Liu \cite{Liu1972}, entropy balance law \eqref{RET:Entropy-BLaw} is regarded as fundamental equation, and balance laws \eqref{RET:BLaws} are treated as constraints. Following the method of multipliers, problem with constraints is transformed into a problem without them by introduction of the co-vector $\mathbf{u}^{\prime}(\mathbf{u})$, so that following compatibility condition is established: 
\begin{equation*}
  \partial_{t} h + \partial_{k} h_{k} - \Sigma 
    = \mathbf{u}^{\prime} \cdot \left( \partial_{t} \mathbf{F} 
    + \partial_{k} \mathbf{F}_{k} - \mathbf{f} \right). 
\end{equation*}
It is easily transformed into: 
\begin{equation} \label{RET:Entropy-Differential}
  \mathrm{d}h = \mathbf{u}^{\prime} \cdot \mathrm{d}\mathbf{F}, \quad 
  \mathrm{d}h_{k} = \mathbf{u}^{\prime} \cdot \mathrm{d}\mathbf{F}_{k}, \quad 
  \Sigma = \mathbf{u}^{\prime} \cdot \mathbf{f} \geq 0. 
\end{equation}

We should make a caesura at this stage, before we proceed with further analysis of the entropy principle. Namely, compatibility conditions \eqref{RET:Entropy-Differential} bring two important novelties. First, once the co-vector of multipliers $\mathbf{u}^{\prime}$ is determined from \eqref{RET:Entropy-Differential}$_{1}$, \eqref{RET:Entropy-Differential}$_{2}$ formally determines the structure of entropy flux. In other words, entropy flux is no longer \emph{a priori} defined, but becomes a constitutive quantity \cite{mueller1967entropy}. Second, residual inequality \eqref{RET:Entropy-Differential}$_{3}$ determines the structure of source terms so that entropy production is semi-definite. This clearly implies that dissipation comes from production terms. 

If we rearrange the compatibility condition \eqref{RET:Entropy-Differential}$_{1}$, the following result emerges: 
\begin{equation*}
  \mathrm{d} \left( \mathbf{u}^{\prime} \cdot \mathbf{F} \right) 
    - \mathbf{u}^{\prime} \cdot \mathrm{d} \mathbf{F} = \mathrm{d}h
  \quad \Rightarrow \quad 
  \mathrm{d} \left( \mathbf{u}^{\prime} \cdot \mathbf{F} - h \right) 
    = \mathbf{u}^{\prime} \cdot \mathrm{d} \mathbf{F}
    =: \mathrm{d} h^{\prime},
\end{equation*}
where $h^{\prime}$ is a new scalar potential, conjugate to $h$. Applying the same procedure to \eqref{RET:Entropy-Differential}$_{2}$, we obtain new potentials by means of Legendre transform of entropy density and entropy flux with respect to densities and fluxes, respectively: 
\begin{equation} \label{RET:Entropy-Potentials}
  h^{\prime} := \mathbf{u}^{\prime} \cdot \mathbf{F} - h, \quad 
  h^{\prime}_{k} := \mathbf{u}^{\prime} \cdot \mathbf{F}_{k} - h_{k}.
\end{equation}
Convexity of the entropy density $h$ with respect to $\mathbf{F}$ allows the transformation $\mathbf{F} \mapsto \mathbf{u}^{\prime}$ to be invertible, not only locally, but globally. Then, from \eqref{RET:Entropy-Potentials} one obtains: 
\begin{equation*}
  \frac{\partial h^{\prime}}{\partial \mathbf{u}^{\prime}} = \mathbf{F}, \quad 
  \frac{\partial h^{\prime}_{k}}{\partial \mathbf{u}^{\prime}} = \mathbf{F}_{k},
\end{equation*}
so that governing equations \eqref{RET:BLaws} can be put into symmetric form: 
\begin{equation} \label{RET:BLawsSymmetric}
  \frac{\partial^{2} h^{\prime}}{\partial \mathbf{u}^{\prime} \partial \mathbf{u}^{\prime}} 
    \frac{\partial \mathbf{u}^{\prime}}{\partial t} 
    + \frac{\partial^{2} h_{k}^{\prime}}{\partial \mathbf{u}^{\prime} 
    \partial \mathbf{u}^{\prime}} 
    \frac{\partial \mathbf{u}^{\prime}}{\partial x_{k}} = \mathbf{f}(\mathbf{u}^{\prime}).
\end{equation} 
Because of the privileged character of multipliers $\mathbf{u}^{\prime}$, that now act as new field variables, they deserved special name \emph{the main field} \cite{ruggeri1981main}, or \emph{entropy variables} \cite{hanouzet2003global}. 

A few remarks may be given at this stage. Liu's procedure \cite{Liu1972} brought a new concept in the study of compatibility between governing equations and entropy balance law. It introduced the multipliers as mediators, but their role went beyond compatibility issues. It appeared that symmetric form of balance laws coincided with Godunov's \cite{godunov1961} analysis of gas dynamics equations, and symmetrization results of Friedrichs and Lax \cite{fried1971systems}. This structure was further exploited in \cite{boillat1997hyperbolic}, and the hierarchy of hyperbolic principal subsystems was recognized. Furthermore, it was shown in the same study that sub-characteristic condition for the characteristic speeds appears as a consequence of sub-entropy law. Finally, hyperbolic systems of balance laws with convex entropy were subject of mathematical analysis of smoothness and global existence of solutions \cite{hanouzet2003global,yong2004entropy}. 

Although the formalism of RET looks rather abstract at first sight, it was applied to numerous particular systems describing different physical processes. From the very beginning, its principal application was related to non-equilibrium thermodynamics, and much of its popularity was gained by remarkable equivalence with Grad's moment method in the kinetic theory of gases \cite{grad1949moments}, especially 13 moments model. It was recognized that in balance laws \eqref{RET:BLaws} densities and fluxes may be endowed with hierarchical structure, with increasing tensorial order of components: 
\begin{equation*}
  \mathbf{F} = 
  \left[
  \begin{array}{c}
  F \\
  F_{i_{1}} \\
  F_{i_{1} i_{2}} \\
  \vdots \\ 
  F_{i_{1} i_{2} \ldots i_{n}} \\
  \end{array}
  \right]; \quad 
  \mathbf{F}_{k} = 
  \left[
  \begin{array}{c}
  F_{k} \\
  F_{i_{1} k} \\
  F_{i_{1} i_{2} k} \\
  \vdots \\ 
  F_{i_{1} i_{2} \ldots i_{n} k} \\
  \end{array}
  \right].
\end{equation*} 
In kinetic theory of gases $F_{i_{1} i_{2} \ldots i_{p}}$ are moments of the distribution function of order $p$. In principle, this hierarchy could be infinite, and in that case it is equivalent to Boltzmann equation. However, when it is truncated at certain order, one must close the system because productions and the last flux are undetermined. Grad expanded the velocity distribution function in terms of generalized Hermite polynomials, and resolved the closure problem by construction of its approximate form. RET, instead, relies on entropy principle and exploits it for the closure problem. 

Relation to the kinetic theory of gases gave RET additional boost, but also put the limits: for a long time, it was restricted to the analysis of monatomic gases, since there was no systematic procedure for derivation of a hyperbolic system of balance laws for polyatomic gases. This obstacle was overcome recently. Starting from the first result \cite{arima2012extended}, a vast amount of results were accumulated and is still growing. An up-to-date collection of main results may be found in \cite{RETpoly}. The crucial distinction with respect to classical monatomic theory is that governing equations are in two hierarchies instead of one. In particular, the model of viscous, heat conducting, polyatomic gas is described by 14 moments, whose governing equations are structured as follows: 
\begin{alignat}{2} \label{RET:14Moments-Equations}
  \partial_{t} F + \partial_{k} F_{k} & = 0; & \quad & 
  \nonumber \\ 
  \partial_{t} F_{i} + \partial_{k} F_{ik} & = 0; & \quad & 
  \\ 
  \partial_{t} F_{ij} + \partial_{k} F_{ijk} & = P_{ij}; & \quad 
    \partial_{t} G_{ll} + \partial_{k} G_{llk} & = 0; 
  \nonumber \\
  & \; & \quad \partial_{t} G_{lli} + \partial_{k} G_{llik} & = Q_{i}. 
  \nonumber
\end{alignat} 
$F-$hierarchy is \emph{momentum-like}, while $G-$hierarchy is \emph{energy-like}, and in physical variables their densities read: 
\begin{alignat}{2} \label{RET:14Moments-Densities}
  F & = \rho; & \quad & 
  \nonumber \\ 
  F_{i} & = \rho v_{i}; & \quad & 
  \\ 
  F_{ij} & = \rho v_{i} v_{j} - t_{ij}; & \quad 
    G_{ll} & = \frac{1}{2} \rho |\mathbf{v}|^{2} + \rho \varepsilon; 
  \nonumber \\
  & \; & \quad G_{lli} & = \left( \frac{1}{2} \rho |\mathbf{v}|^{2} + \rho \varepsilon \right)
    v_{i} - t_{ij} v_{j} + q_{i},  
  \nonumber
\end{alignat} 
where $t_{ij} = - (p + \Pi) \delta_{ij} + \sigma_{\langle ij \rangle}$ is the stress tensor, $p$ is the hydrostatic pressure, $\Pi$ is the dynamic pressure, and $\sigma_{\langle ij \rangle}$ is the traceless viscous stress, while $q_{i}$ is the heat flux. In such a way, the last equation of $F-$hierarchy in \eqref{RET:14Moments-Equations} becomes balance law for stress tensor, while the last equation of $G-$hierarchy becomes balance law for heat flux. 

It must be emphasized that, except in the simplest cases of 5 or 6 moments equations, closure obtained by means of Liu's method of multipliers can be achieved only in the neighborhood of local equilibrium. In that sense it is not exact, but approximate. As a consequence, productions $P_{ij}$ and $Q_{i}$, and entropy density are determined only approximately. In particular, we have the productions: 
\begin{equation} \label{RET:14Moments-Productions}
  P_{ij} \approx \frac{1}{\tau_{S}} \sigma_{\langle ij \rangle} 
    - \frac{1}{\tau_{\Pi}} \Pi \delta_{ij}; 
  \quad 
  Q_{i} \approx v_{k} P_{ik} - \frac{1}{\tau_{q}} q_{i},   
\end{equation}
and the entropy density and entropy flux: 
\begin{align} \label{RET:14Moments-Entropy}
  h & \approx \rho s - \frac{3 D}{4 (D-3) p T} \Pi^{2} 
    - \frac{1}{4 p T} \sigma_{\langle ij \rangle} \sigma_{\langle ij \rangle} 
    - \frac{\rho}{(D+2) p^{2} T} q_{i} q_{i}; 
  \\
  h_{k} & \approx h v_{k} + \frac{q_{k}}{T} - \frac{2}{(D+2) p T} \Pi q_{k} 
    + \frac{2}{(D+2) p T} q_{i} \sigma_{\langle ik \rangle}, 
  \nonumber
\end{align} 
where $s$ is specific entropy in equilibrium. In \eqref{RET:14Moments-Productions}, $\tau_{S}$, $\tau_{\Pi}$ and $\tau_{q}$ are relaxation times, that are functions of objective quantities, typically mass density $\rho$ and temperature $T$. 

Here we face important limitations in the practical application of described closure procedure. First, higher-order theories admit only approximate closure, thus restricting the application to processes in a certain neighborhood of the local equilibrium. This is also manifested through the loss of hyperbolicity in certain regions of state space. That was the subject of \cite{brini2001hyper}. Second, production terms are determined up to scalar positive functions, relaxation times, that are related to transport coefficients. This is a typical limitation of macroscopic theory, and cannot be overcome within its framework. To do that, one must step outside, either towards some more refined theoretical approach, or towards experimental data. 

\subsection{Molecular extended thermodynamics---application of the maximum entropy principle} 

While RET successfully combined the methodology of continuum mechanics, i.e., application of the entropy principle, with the hierarchical structure of balance laws resembling the moment method of Grad, molecular extended thermodynamics (also termed extended thermodynamics of moments), relied on the application of Maximum Entropy Principle (MEP). MEP is a generalization of the well-known result of classical thermodynamics that entropy reaches a maximum in equilibrium. 

In the kinetic theory of gases, state of the system is described by velocity distribution function $f(t, \mathbf{x}, \boldsymbol{\xi})$, where $\boldsymbol{\xi} \in \mathbb{R}^{3}$ is the molecular velocity. Its evolution is determined by the Boltzmann equation: 
\begin{equation*}
  \partial_{t} f + \xi_{k} \partial_{k} f = Q(f,f), 
\end{equation*}
where $Q(f,f)$ is integral collision operator that describes the mechanism of interaction between the particles. In this context MEP acquires variational character. Namely, it assumes that the actual state of the system is monitored through macroscopic quantities determined as moments of $f$. Therefore, for the kinetic entropy defined as functional $h = h[f]$, the actual distribution function is the one that maximizes $h$, subject to constraints which define macroscopic variables as moments of $f$. 

The solution of the variational problem depends on the choice of macroscopic variables. If the constraints are to be $\rho$, $\rho \mathbf{v}$, and $\rho \varepsilon$, one obtains the Maxwellian equilibrium distribution function as a maximizer. Kogan \cite{kogan-book} showed that extending the list of constraints by the stress tensor and heat flux leads to Grad's distribution function as a maximizer. This gave a stimulus to further application of MEP and led to a striking conclusion that \emph{exact} solution of the variational problem formulated above is equivalent to the results of extended thermodynamics obtained by the procedure described in the previous subsection. Here, an explanation should be given for the word \emph{exact}. Namely, the solution to the variational problem is not particularly challenging---it is an exponential function of the polynomial of molecular velocities multiplied by appropriate Lagrange multipliers. However, computation of Lagrange multipliers 
 in terms of macroscopic variables could be problematic if the largest degree of the moment is odd---in that case, the moments are diverging. That was observed already by Kogan, and it was further analyzed by Dreyer \cite{dreyer1987maximisation}, leading to the approximation of the exact solution in the neighborhood of Maxwellian distribution. That was the way in which the closure problem was resolved in the case of odd highest moments. More mathematically intoned formulation of the MEP was given by Levermore \cite{levermore1996moment}. 

Although MEP may be regarded as just another route to the hyperbolic system of balance laws already given by RET, there is an interest in its development. It may fill the gap between the endpoint reached by RET and the fully closed system of balance laws. An illustrative example will be the hyperbolic system of balance laws for polyatomic gases. 

Starting from the statistical mechanical considerations \cite{borgnakke1975statistical}, it was shown \cite{bourgat1994micro,desvillettes2005kinetic} that one may use a single additional continuous variable $I$ to describe the internal degrees of freedom of molecules in polyatomic gas, and to recover the internal energy density of polytropic gases. Within this kinetic framework, it was formulated a MEP for polyatomic gases \cite{pavic2013maximum} that recovered the results \eqref{RET:14Moments-Equations}-\eqref{RET:14Moments-Entropy} of RET. More precisely, for the kinetic entropy defined as: 
\begin{equation} \label{RET:MEP-Entropy}
  h[f] = - k_{\mathrm{B}} \int_{\mathbb{R}^{3}} \int_{0}^{\infty} f \log f \varphi(I) 
    \mathrm{d}I \mathrm{d}\boldsymbol{\xi},
\end{equation}
one searches for the maximizer $f(t,\mathbf{x},\boldsymbol{\xi},I)$ subject to constraints: 
\begin{align} \label{RET:MEP-Constraints}
  \left[
  \begin{array}{c}
    \rho \\ 
    0_{i} \\ 
    - t_{ij} \\
  \end{array}
  \right]
  & = \int_{\mathbb{R}^{3}} \int_{0}^{\infty} m
  \left[
  \begin{array}{c}
    1 \\ 
    C_{i} \\ 
    C_{i} C_{j} \\
  \end{array}
  \right]
  f \varphi(I) \mathrm{d}I \mathrm{d}\mathbf{C}, 
  \\
  \left[
  \begin{array}{c}
    \rho \varepsilon \\ 
    q_{i} \\ 
  \end{array}
  \right]
  & = \int_{\mathbb{R}^{3}} \int_{0}^{\infty} m
  \left[
  \begin{array}{c}
  \frac{1}{2} |\mathbf{C}|^{2} + \frac{I}{m} \\ 
  \left(\frac{1}{2} |\mathbf{C}|^{2} + \frac{I}{m}\right) C_{i} \\ 
  \end{array}
  \right]
  f \varphi(I) \mathrm{d}I \mathrm{d}\mathbf{C} 
  \nonumber
\end{align}
where $\mathbf{C} = \boldsymbol{\xi} - \mathbf{v}$ is the peculiar velocity. A non-negative measure $\varphi(I) \mathrm{d}I$ is property of the model. By taking $\varphi(I) = I^{\alpha}$, for $\alpha > -1$, one obtains the following 14 moments velocity distribution function: 
\begin{align} \label{RET:MEP-f14}
  f \approx f_{14} & = f_{\mathrm{E}} \left\{ 1 - \frac{\rho}{p^{2}} q_{i} C_{i} 
    + \frac{\rho}{2 p^{2}} \left[ \left( \frac{5}{2} + \alpha \right)(1 + \alpha)^{-1} 
    \Pi \delta_{ij} - \sigma_{\langle ij \rangle} \right] C_{i} C_{j} \right.
  \\
  & \quad \left. - \frac{3}{2(1 + \alpha)} \frac{\rho}{p^{2}} \Pi 
    \left( \frac{1}{2} |\mathbf{C}|^{2} + \frac{I}{m} \right) 
    + \left( \frac{7}{2} + \alpha \right)^{-1} \frac{\rho^{2}}{p^{3}} q_{i}
    \left( \frac{1}{2} |\mathbf{C}|^{2} + \frac{I}{m} \right) C_{i} \right\}, 
  \nonumber 
\end{align}
where $f_{\mathrm{E}}$ and $q(T)$: 
\begin{equation*}
  f_{\mathrm{E}} = \frac{\rho}{m q(t)} \left( \frac{m}{2 \pi k_{\mathrm{B}} T} \right)^{3/2} 
    \exp \left\{ - \frac{1}{k_{\mathrm{B}} T} 
    \left( \frac{1}{2} |\mathbf{C}|^{2} + \frac{I}{m} \right) \right\}, 
  \quad 
  q(T) = \int_{0}^{\infty} \exp \left( - \frac{I}{k_{\mathrm{B}} T} \right) 
    \varphi(I) \mathrm{d}I,
\end{equation*}
are equilibrium distribution and auxiliary function, respectively. Densities \eqref{RET:14Moments-Densities} are recovered as moments of $f_{14}$ through \eqref{RET:MEP-Constraints}, as well as entropy and entropy flux \eqref{RET:14Moments-Entropy} using the relation $\alpha = (D - 5)/3$. However, the main advantage comes from computation of the productions \eqref{RET:14Moments-Productions} as moments, i.e. weak form of the collision operator: 
\begin{equation*}
  P_{ij} = \int_{\mathbb{R}^{3}} \int_{0}^{\infty} m \xi_{i} \xi_{j} Q(f,f) 
    \varphi(I) \mathrm{d}I \mathrm{d}\boldsymbol{\xi}, \quad 
  Q_{i} = \int_{\mathbb{R}^{3}} \int_{0}^{\infty} m \left( \frac{1}{2} |\boldsymbol{\xi}|^{2} 
    + \frac{I}{m} \right) \xi_{i} Q(f,f) \varphi(I) \mathrm{d}I \mathrm{d}\boldsymbol{\xi}. 
\end{equation*} 
In \cite{pavic2014moment} it was used collision cross section that describes variables hard spheres, with an additional parameter $s > -3/2$, so that relaxation times were explicitly computed: 
\begin{align}
  \frac{1}{\tau_{S}} & = K \frac{2^{2s+4}}{15} 
    \frac{\rho (k_{\mathrm{B}}T)^{2}}{m q(T)^{2}} \sqrt{\pi} 
    \left( \frac{k_{\mathrm{B}}T}{m} \right)^{s} \Gamma\left[ s + \frac{3}{2} \right], 
  \nonumber \\ 
  \frac{1}{\tau_{\Pi}} & = K \left( \frac{5}{2} + \alpha \right) (1 + \alpha)^{-1}  
    \frac{2^{2s+6}}{3(2s+5)(2s+7)} 
    \frac{\rho (k_{\mathrm{B}}T)^{2}}{m q(T)^{2}} \sqrt{\pi} 
    \left( \frac{k_{\mathrm{B}}T}{m} \right)^{s} \Gamma\left[ s + \frac{3}{2} \right], 
  \\ 
  \frac{1}{\tau_{q}} & = K \left( \frac{7}{2} + \alpha \right)^{-1}  
    \frac{2^{2s+5}(s(2s+15)+30)}{9(2s+5)(2s+7)} 
    \frac{\rho (k_{\mathrm{B}}T)^{2}}{m q(T)^{2}} \sqrt{\pi} 
    \left( \frac{k_{\mathrm{B}}T}{m} \right)^{s} \Gamma\left[ s + \frac{3}{2} \right]. 
  \nonumber
\end{align}
This result was used to adapt the parameter $s$ so that model yields physically correct value of Prandtl number. In such a way, closure of 14 moments model is completed by means of simultaneous application of RET and MEP. 

We would like to draw attention to the question of non-linearity in the context of RET, especially related to production terms. To that end, we mention two examples. The first one is the multi-temperature mixture of Euler fluids. The model was established within the framework of RET in \cite{ruggeri2007hyperbolic}. Closure obtained in this case was \emph{exact}. Nevertheless, there remained a gap in the sense of phenomenological positive semi-definite matrices. More precise characterization was impossible within the macroscopic approach. Only in the special case of binary mixture it was determined the explicit form of the source terms \cite{ruggeri2009average,madjarevic2013shock,madjarevic2014shock}. Recently \cite{pavic2019multi}, using the kinetic theory approach, it was determined the non-linear structure of source terms in the multi-temperature mixture. Since they were structurally different from those obtained by RET, compatibility was achieved in a linear approximation 
 of both. In such a way, phenomenological coefficients of RET were again determined in conjunction with the kinetic theory approach. The second example is concerned with non-linear closure using MEP by means of higher degree of multipliers used for expansion in the neighborhood of Maxwellian distribution. This leads to non-linear closure for non-convective fluxes, as shown in \cite{brini2002entropy,brini2020second}.

\subsection{Note on boundary conditions} 

Derivation of appropriate boundary conditions is a tremendous open problem in RET. There is no widely accepted approach, and this note will provide some of the standpoints and procedures that appeared in the literature. 

One possible approach is developed and exploited in the works of Struchtrup and Torrilhon \cite{struchtrup2005macroscopic,struchtrup2007regularization}. The main features that distinguish this approach will be briefly explained. First, analysis is built up around so-called R13 equations---parabolic regularization of Grad's 13 moments equations obtained by adding the terms of Super-Burnett order. Construction of the model may be justified as regularization in the neighborhood of pseudo-equilibrium manifold, determined by Grad's velocity distribution, just as NSF model presents regularization in the vicinity of equilibrium manifold determined by the Maxwellian \cite{struchtrup2005macroscopic}. To determine the unknown non-convective fluxes for higher-order equations, the authors proceed in a classical way and exploit the entropy production rate. The outcome is non-local constitutive relations that bring regularizing terms in the model. 

Second, the key for the derivation of boundary conditions is the entropy production rate due to collisions of the atoms with a solid wall. It is determined from the linear form of the entropy inequality, and the entropy flux at the wall. According to \cite{struchtrup2007regularization}, since the entropy production rate must be non-negative, entropy flux out of the gas wall interface must be larger than the entropy flux into the interface. This implies the bilinear structure of the entropy production rate and determines the fluxes at the boundary in terms of their driving forces. Such boundary conditions generalize the Maxwell-Smoluchowski velocity slip-temperature jump conditions. 

Another approach is proposed by Barbera et al. \cite{barbera2004determination} in the context of heat transfer problems in rarefied gases. It was motivated by the observation that rarefied gases, and the moments which describe its state, are subject to thermal fluctuations. They are so rapid that gas cannot adjust its fields to the actual values of non-controllable data neither in the bulk region, nor at the boundary. Therefore, it is assumed that the gas adjusts its field variables to the mean values of fluctuating data. Thermal fluctuations are determined by the entropy evaluated for Grad's approximation of the velocity distribution function. 

These two examples demonstrate that the derivation of boundary data for higher-order moments in RET is a demanding task, which cannot be resolved in a universal manner. In both approaches mentioned above, boundary conditions are consequences of some general physical principle/insight. However, we still need a comprehensive comparative analysis of different approaches. 

\section{Study closure}

Since the seminal paper of Coleman and Noll \cite{Coleman1963}, the second law of thermodynamics took the role of constraint on the admissible forms of constitutive relations. This introduced the notion of \emph{thermodynamic consistency} for the continuum models. Nevertheless, certain questions still remain open, and our aim was to enlighten some mathematical aspects of continuum theories related to the models of non-equilibrium/irreversible processes. 

Throughout the study, one may follow the evolution of the idea proposed by Coleman and Noll. Within the framework of CIT (Section 2), after the discussion of balance laws and the notion of entropy, the second law was exploited in the analysis of heat conduction in rigid bodies. Apart from classical result---linear relation between the heat flux and temperature gradient---a non-linear closure was proposed along with an inherent problem of determination of the phenomenological coefficients. This line of analysis was pursued in NET-IV (Section 3) in which internal variables were introduced, and generalizations of the heat conduction problem were discussed. This led to the Maxwell-Cattaneo-Vernotte, and, after the generalization of entropy flux, to the Guyer-Krumhansl equation. The study in this framework was then expanded to the problems with coupled mass and heat transport, which naturally motivated comparison with EIT and RET. Finally, within RET (Section 4), it was presented 
 the formalism that yields the governing equations, which predict finite speeds of propagation of disturbances. The list of field variables is extended, the structure of equations is strict, and the entropy balance law yields a closure for non-convective fluxes and source terms. Although the entropy balance law retained the central role, this r\'{e}sum\'{e} demonstrates that the introduction of new assumptions could enrich the picture of non-equilibrium/irreversible processes. The way in which it is achieved depends on the aims and restrictions of a certain approach. For example, NET-IV relies on internal variables and (extended) entropy balance law to produce appropriate constitutive relations, which eventually get the rate type form. On the other hand, RET imposes governing equations in the form of a hyperbolic system of balance laws for an extended set of variables. To determine the phenomenological coefficients, NET-IV relies on experimental data, whereas RET may use the 
 possibility of looking at first principles, i.e., kinetic theory of gases, to close the system. 

Another issue that may be followed is the problem of initial and boundary conditions. This is a problem that has two equally important sides---mathematical and physical. Boundary conditions should be formulated such that the well-posedness of the problem is preserved. At the same time, the data for initial and boundary conditions are supposed to be measured or controlled, and this may cause the problem in generalized approaches. Our review shows that within CIT, it is a classical and well-understood problem, both mathematically and experimentally. NET-IV proposes a generalization of convection-type boundary data, based upon constitutive relation, in a manner that is analogous to CIT. However, RET, which is primarily concerned with the behavior of rarefied gases, grabs for physical arguments that may yield appropriate information about the conditions on the solid boundary. From this perspective, it becomes apparent that there is no unifying approach to the problem of boundary 
 conditions. It is still an open and challenging field for study. 

Lastly, there is one more question which will remain unanswered, perhaps because it may be related to taste or philosophical standpoint. As mentioned in analysis of NET-IV, equations obtained in this way may be regarded as parabolic envelope to corresponding (subordinate) hyperbolic system. On the other hand, it is a standard procedure in RET to recover the parabolic (thermodynamic) limit by asymptotic methods in the small relaxation time limit \cite{ruggeri2012can} (even R13 equations may be obtained in this way starting from higher order hyperbolic system). It is really a matter of one's preferences which system will be regarded as ``fundamental'', and which one will be treated as its approximation.

As an outlook, we also mention the framework of GENERIC (General Equation for the Non-Equilibrium Reversible--Irreversible Coupling) \cite{Grmela1997,Ottinger1997,Ottinger2005,Pavelka2018}, which can be interpreted as a thermodynamically consistent generalization of Hamiltonian mechanics, and as such, it raises additional mathematical questions, such as the geometric structure behind thermodynamics. GENERIC derives the reversible and irreversible contributions separately from each other, which looks artificial in some cases, such as Fourier heat conduction as a purely dissipative process, but in a more general and complex situation, it is a productive method for deriving dynamical models. That sort of separation can be useful in the construction of specific numerical methods, see, e.g., \cite{Shang2020,Pavelka2019,Portillo2017,Betsch2019}. The comparison of GENERIC with other non-equilibrium thermodynamical frameworks is currently under investigation  \cite{Szucs2019, Ottinger2020}.


\vspace{6pt} 

{\section{Acknowledgement} The authors are thankful to Péter Ván for the valuable discussions.}


The research was funded (M. Szücs and R. Kovács) by the grants of National Research, Development and Innovation Fund - NKFIH 130378. The research reported in this paper was supported by the BME NC TKP2020 grant of NKFIH Hungary; and (S. Simi\'c) by the Ministry of Education, Science and Technological Development of the Republic of Serbia (Grant No. 451-03-68/2020-14/200125).

\end{document}